\pdfoutput=1
\documentclass[prb,twocolumn,aps]{revtex4}
\usepackage{graphicx,graphics,color,epsfig}
\usepackage{bm}
\usepackage{amsmath}
\usepackage{amssymb}
\usepackage{epstopdf}

\makeatletter

\newcommand{\Rmnum}[1]{\expandafter\@slowromancap\romannumeral #1@}
\newcommand*{\rom}[1]{\expandafter\@slowromancap\romannumeral #1@}
\makeatother

\usepackage{color}

\begin{document}

\title{In-gap bound states induced by a single nonmagnetic impurity in sign-preserving $s$-wave superconductors with incipient bands}

\author{Yi Gao,$^{1}$ Yan Yu,$^{1}$ Tao Zhou,$^{2}$ Huaixiang Huang,$^{3}$ and Qiang-Hua Wang$^{4,5}$}
\affiliation{$^{1}$Department of Physics, Nanjing Normal University, Nanjing, 210023, China\\
$^{2}$College of Science, Nanjing University of Aeronautics and Astronautics, Nanjing, 210016, China\\
$^{3}$Department of Physics, Shanghai University, Shanghai, 200444, China\\
$^{4}$National Laboratory of Solid State Microstructures $\&$ School of Physics, Nanjing
University, Nanjing, 210093, China\\
$^{5}$Collaborative Innovation Center of Advanced Microstructures, Nanjing 210093, China}

\begin{abstract}
We have investigated the in-gap bound states (IGBS) induced by a single nonmagnetic impurity in multiband superconductors with incipient bands. Contrary to the naive expectation, we found that even if the superconducting (SC) order parameter is sign-preserving $s$-wave on the Fermi surfaces, the incipient bands may still affect the appearance and locations of the IGBS, although the gap between the incipient bands and the Fermi level is much larger than the SC gap. Therefore in scanning tunneling microscopy experiments, the IGBS induced by a single nonmagnetic impurity are not the definitive evidences for the sign-changing order parameter on the Fermi surfaces. Our findings have special implications for the experimental determination of the pairing symmetry in the FeSe-based superconductors.
\end{abstract}


\maketitle

The recent discovery of high-T$_c$ superconductivity in some FeSe-based superconductors, such as Li$_{1-x}$Fe$_x$OHFe$_{1-y}$Se,\cite{chenxh,Johrendt,zhaozx} Li$_x$(NH$_2$)$_y$(NH$_3$)$_{1-y}$Fe$_2$Se$_2$ \cite{clarke} and A$_x$Fe$_{2-y}$Se$_2$ (A=Rb, Cs, K),\cite{chenxl,chenxh2,conder} as well as monolayer FeSe grown on SrTiO$_{3}$,\cite{xueqk1} has attracted great interest among the condensed matter community. These materials have only electron Fermi surfaces around the Brillouin zone (BZ) corner, while the hole Fermi surfaces around the BZ center, which commonly exist in the usual iron pnictide superconductors, disappear since the hole bands sink below the Fermi level and become incipient.\cite{zhouxj1,zhouxj2,fengdl1,fengdl2,shenzx1,zhouxj3,fengdl4,ding,zhouxj4,fengdl5,fengdl6,shenzx2}

The superconducting (SC) mechanism and pairing symmetry in these materials are currently under hot debate. Theoretically it was suggested that, due to the disappearance of the hole Fermi surfaces, the spin fluctuation between the electron Fermi surfaces results in a nodeless $d$-wave pairing symmetry,\cite{aoki,scalapino,leedh,balatsky} where the SC order parameter changes sign among the Fermi surfaces. However this may lead to gap nodes or extreme minima in the 2Fe/cell BZ.\cite{mazin2} In contrast, $s$-wave pairing symmetry has also been predicted with sign-preseving order parameter among the Fermi surfaces.\cite{zhou,hu,kontani,wang} Sign-changing $s$-wave pairing symmetry has been predicted as well, where the order parameter changes sign between the inner and outer Fermi surfaces.\cite{mazin2} Experimentally, angle-resolved photoemission spectroscopy found that the SC gap magnitude along the Fermi surfaces shows no apparent nodes or extreme minima,\cite{zhouxj1,shenzx1,zhouxj3,fengdl4,zhouxj4,fengdl5,fengdl6,shenzx2} therefore the nodeless $d$-wave case is less supported. A spin resonance observed by inelastic neutron scattering (INS) \cite{Boothroyd,zhaoj1,Boothroyd1,Inosov1,Inosov2,Boothroyd2,Inosov3,zhaoj} was interpreted as the possible sign change of the order parameter among the Fermi surfaces. Scanning tunneling microscopy (STM) results are controversial since Refs. \onlinecite{fengdl3} and \onlinecite{fengdl7} claim no such sign change while Ref. \onlinecite{wen} reaches the opposite conclusion.

Up to now, most of the studies in the FeSe-based superconductors have focused on the sign of the order parameter on the Fermi surfaces, with little attention paid to the SC pairing on the incipient bands which do not form Fermi surfaces. In LiFe$_{1-x}$Co$_x$As, a $\sim5$meV SC gap has been observed on a hole band which sinks $\sim8$meV below the Fermi level \cite{ding2} while in FeTe$_{0.6}$Se$_{0.4}$, a $\sim1.1$meV gap is found on an electron band lying $\sim0.7$meV above the Fermi level.\cite{shin} However in the FeSe-based superconductors, the hole bands sink $\sim100$meV below the Fermi level and whether they are SC or not is unclear. For example, Refs. \onlinecite{hu} and \onlinecite{wang} suggest that the incipient bands are SC and the sign of the order parameter on them is opposite to that on the Fermi surfaces, while Ref. \onlinecite{kontani} states that the sign of the order parameter on the incipient bands is the same as that on the Fermi surfaces. In addition, Ref. \onlinecite{bang2} claims no SC pairing on the incipient bands. Furthermore, whether these incipient bands will affect the experimental interpretation of the sign of the order parameter on the Fermi surfaces is less explored. In Ref. \onlinecite{gao1}, it was shown that the spin susceptibility is not affected by the incipient bands qualitatively, therefore INS is reliable in detecting the sign of the order parameter on the Fermi surfaces. Another commonly accepted criterion is the in-gap bound states (IGBS) induced by a nonmagnetic impurity, which can be measured by STM. If such states exist, then it is believed that the order parameter should change sign on the Fermi surfaces. Otherwise the order parameter should preserve its sign on the Fermi surfaces if the IGBS do not exist. Such properties are successful in identifying the unconventional nature of different classes of superconductors.\cite{zhujx,dgzhang,kontani2,hhwen,pennec} In the FeSe-based superconductors, Ref. \onlinecite{zhujx2} theoretically proposed that the IGBS can exist only if the order parameter changes sign on the Fermi surfaces in the full gap opening case. In addition it claims that it is not important whether there is a SC gap on the incipient bands. Therefore Refs. \onlinecite{fengdl3} and \onlinecite{fengdl7} claim sign-preserving $s$-wave pairing since they observe no IGBS while Ref. \onlinecite{wen} suggests sign-changing order parameter on the Fermi surfaces since the observation of the IGBS.

In this work, we show that, when the incipient bands are present, IGBS can be induced by a single nonmagnetic impurity, even if the order parameter is sign-preserving $s$-wave on the Fermi surfaces. The locations of the IGBS can vary from the gap edges to deep inside the gap, depending on the scattering strength of the impurity, the density of states (DOS) on the Fermi level, the relative sign between the order parameter on the incipient bands and that on those bands crossing the Fermi level, as well as the details of the band structure. Therefore, special caution has to be taken when inferring the sign of the order parameter on the Fermi surfaces based on the STM observation of the IGBS induced by a single nonmagnetic impurity.

\begin{figure}
\includegraphics[width=1\linewidth]{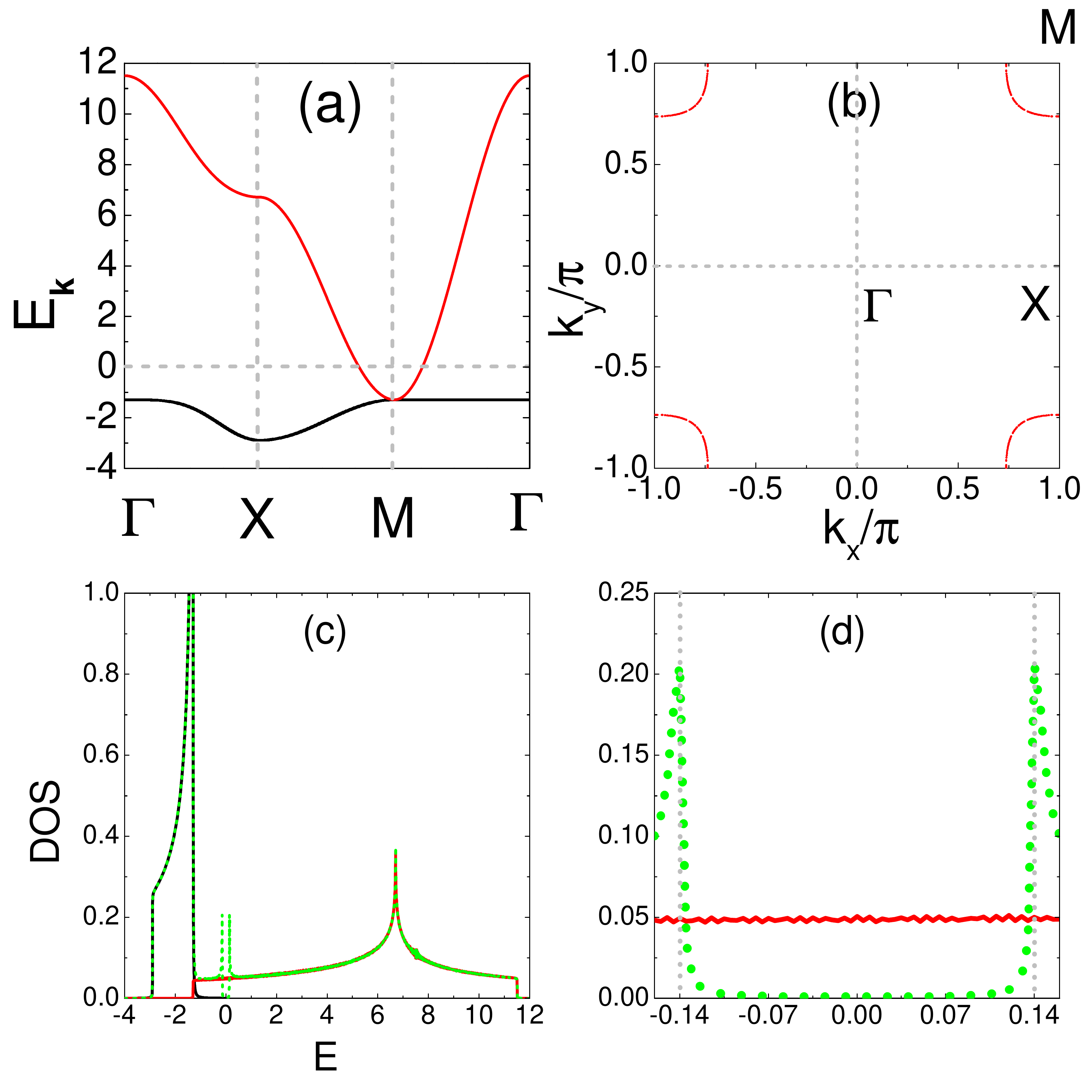}
 \caption{\label{band} (color online) (a) Band structure of the incipient (black) and active (red) bands along the high-symmetry directions. The Fermi level is denoted by the gray dotted line at $E_{\mathbf{k}}=0$. (b) Fermi surfaces in the 2Fe/cell BZ. (c) DOS of the incipient (black solid) and active (red solid) bands, as well as that in the SC state (green dashed). (d) is the DOS close to the Fermi level shown in (c), where the two gray dotted lines indicate the SC coherence peaks at $\pm\Delta$. The DOS in the SC state in (c) and (d) is calculated by setting $\Delta_1=0$.}
\end{figure}

In the following, we adopt a 2D tight-binding model similar to that proposed in Ref. \onlinecite{gaoy_fese}, where each unit cell contains two inequivalent sublattices $A$ and $B$. The coordinate of the sublattice $A$ in the unit cell $(i,j)$ is $\mathbf{R}_{ij}=(i,j)$ while that for the sublattice $B$ is $\mathbf{R}_{ij}+\mathbf{d}$, with $\mathbf{d}$ being $(0.5,0.5)$. For illustrative purpose, we simply consider only one orbital on each sublattice. The Hamiltonian can be written as $H=\sum_{\mathbf{k}}\psi_{\mathbf{k}}^{\dag}A_{\mathbf{k}}\psi_{\mathbf{k}}$, where $\psi_{\mathbf{k}}^{\dag}=(c_{\mathbf{k}A\uparrow}^{\dag},c_{\mathbf{k}B\uparrow}^{\dag},c_{-\mathbf{k}A\downarrow},c_{-\mathbf{k}B\downarrow})$ and
\begin{eqnarray}
\label{h}
A_{\mathbf{k}}&=&\begin{pmatrix}
M_{\mathbf{k}}&D_{\mathbf{k}}\\D_{\mathbf{k}}^{\dag}&-M_{-\mathbf{k}}^{T}
\end{pmatrix},
M_{\mathbf{k}}=\begin{pmatrix}
\epsilon_{A\mathbf{k}}&\epsilon_{T\mathbf{k}}\\
\epsilon_{T\mathbf{k}}^{*}&\epsilon_{B\mathbf{k}}
\end{pmatrix}.
\end{eqnarray}
Here $c_{\mathbf{k}A\uparrow}^{\dag}/c_{\mathbf{k}B\uparrow}^{\dag}$ creates a spin up electron with momentum $\mathbf{k}$ and on the $A/B$ sublattice. $\epsilon_{A\mathbf{k}}=-2(t_{2}\cos k_{x}+t_{3}\cos k_{y})-\mu$, $\epsilon_{B\mathbf{k}}=-2(t_{2}\cos k_{y}+t_{3}\cos k_{x})-\mu$ and $\epsilon_{T\mathbf{k}}=-t_{1}[1+e^{-ik_x}+e^{-ik_y}+e^{-i(k_x+k_y)}]$. In addition, $M_{\mathbf{k}}$ and $D_{\mathbf{k}}$ are the tight-binding and pairing parts of the system, respectively. Throughout this work, $\mathbf{k}$ is defined in the 2Fe/cell BZ and the energies are in units of 0.1eV. In the following we set $t_{1-3}=1.6,0.4,-2$ and $\mu=-1.9$. The band structure as well as the pairing function in the band basis can be obtained through a unitary transform as
\begin{eqnarray}
\label{unitary}
Q_{\mathbf{k}}^{\dag}M_{\mathbf{k}}Q_{\mathbf{k}}&=&\begin{pmatrix}
E_{\mathbf{k}1}&0\\
0&E_{\mathbf{k}2}
\end{pmatrix},\nonumber\\
\Delta_{\mathbf{k}}&=&Q_{\mathbf{k}}^{\dag}D_{\mathbf{k}}Q_{-\mathbf{k}}^{*}=\begin{pmatrix}
\Delta_1&0\\
0&\Delta_2
\end{pmatrix}.
\end{eqnarray}
Here $E_{\mathbf{k}1}$ ($E_{\mathbf{k}2}$) is the energy of the incipient (active) band ($E_{\mathbf{k}1}\leq E_{\mathbf{k}2}$), whose dispersion along the high-symmetry directions is shown in Fig. \ref{band}(a). The top of the incipient band and the bottom of the active band are both located at $E_g\approx100$meV below the Fermi level while only the active band forms Fermi surfaces, as shown in Fig. \ref{band}(b). The diagonal (off-diagonal) components of $\Delta_{\mathbf{k}}$ represent the intraband (interband) pairings. We assume only intraband pairing and $|\Delta_1|,|\Delta_2|\ll E_g$, with $\Delta_1$ ($\Delta_2$) being the pairing function on the incipient (active) band. The pairing function is supposed to be momentum independent for simplicity, so the pairing order parameter is sign-preserving $s$-wave on the Fermi surfaces. Figures \ref{band}(c) and \ref{band}(d) show the DOS of the incipient and active bands, as well as that in the SC state. We can see that the DOS of the active band is close to a constant in the vicinity of the Fermi level and since only the active band forms Fermi surfaces, two SC coherence peaks are located at $\pm\Delta=\pm\Delta_2$. In this work, we set $\Delta_2=\Delta=0.14$ ($\sim14$meV), unless otherwise stated.

When a single nonmagnetic impurity is placed at the sublattice $A$ in the unit cell $(0,0)$, it plays
the role of an onsite potential scatter and the impurity Hamiltonian can be written as $H_{imp}=V\sum_{\sigma=\uparrow,\downarrow}c_{(0,0)A\sigma}^{\dag}c_{(0,0)A\sigma}$. Following the treatment for classical spins in superconductors by H. Shiba,\cite{shiba} the $T$-matrix can be expressed as $T(\omega)=[I-U_sG_{0}(\omega)]^{-1}\frac{U_s}{N}$,
where $G_{0}(\omega)=\frac{1}{N}\sum_k g_{0}(\mathbf{k},\omega)$ and $g_{0}(\mathbf{k},\omega)=(\omega I-A_{\mathbf{k}})^{-1}$. Here $I$ is a $4\times4$ unit matrix, $N$ is the number of the unit cells and the nonzero elements of $U_s$ are $U_s^{11}=-U_s^{33}=V$. In fully gapped superconductors, the poles of $T(\omega)$ at $|\frac{\omega}{\Delta}|<1$ signify the impurity-induced IGBS,\cite{zhujx} which should show up when $p(\omega)=det[I-U_sG_{0}(\omega)]=0$, where $p(\omega)=1-V(G_0^{11}-G_0^{33})-V^2(G_0^{11}G_0^{33}-G_0^{13}G_0^{31})$ in the present model. Here
\begin{eqnarray}
\label{G0}
G_0^{11}(\omega)&=&-G_0^{33}(-\omega)=\frac{1}{N}\sum_{i=1}^{2}\sum_k\frac{|Q_\mathbf{k}^{1i}|^2(\omega+ E_{\mathbf{k}i})}{\omega^2- \xi_{\mathbf{k}i}^{2}},\nonumber\\
G_0^{13}(\omega)&=&G_0^{31}(\omega)=\frac{1}{N}\sum_{i=1}^{2}\sum_k\frac{|Q_\mathbf{k}^{1i}|^2\Delta_i}{\omega^2-\xi_{\mathbf{k}i}^{2}},
\end{eqnarray}
with $\xi_{\mathbf{k}i}=\sqrt{E_{\mathbf{k}i}^2+\Delta_i^2}$. Since the $A$ and $B$ sublattices are related by exchanging $k_x$ and $k_y$, while the bands are also symmetric with respect to exchanging $k_x$ and $k_y$, therefore $|Q_\mathbf{k}^{1i}|^2$ in Eq. (\ref{G0}) can be replaced by $\frac{1}{2}$. In the following, we discuss the condition for $p(\omega)=0$ at $|\frac{\omega}{\Delta}|<1$.

Since $|\omega|<|\Delta|\ll|E_{\mathbf{k}1}|$, we have
\begin{eqnarray}
\label{eta}
\frac{1}{N}\sum_k\frac{\omega(\Delta_1)}{\omega^2-\xi_{\mathbf{k}1}^{2}}&=&-\frac{1}{N}\sum_k\frac{\omega(\Delta_1)}{\xi_{\mathbf{k}1}^{2}}(1-\frac{\omega^2}{\xi_{\mathbf{k}1}^{2}})^{-1}\nonumber\\
&\approx&-\frac{\omega(\Delta_1)}{N}\sum_k\frac{1}{\xi_{\mathbf{k}1}^{2}}\nonumber\\
&=&-\langle\frac{1}{\xi_{\mathbf{k}1}^{2}}\rangle\omega(\Delta_1)=\eta\omega(\Delta_1).
\end{eqnarray}
Here $\langle\cdots\rangle$ represents the average over the BZ. Furthermore,
\begin{eqnarray}
\label{delta}
\frac{1}{N}\sum_k\frac{E_{\mathbf{k}1}}{\omega^2-\xi_{\mathbf{k}1}^{2}}&=&-\frac{1}{N}\sum_k\frac{E_{\mathbf{k}1}}{\xi_{\mathbf{k}1}^{2}}(1-\frac{\omega^2}{\xi_{\mathbf{k}1}^{2}})^{-1}\nonumber\\
&\approx&-\langle\frac{E_{\mathbf{k}1}}{\xi_{\mathbf{k}1}^2}\rangle=\delta,
\end{eqnarray}
and
\begin{eqnarray}
\label{gamma}
\frac{1}{N}\sum_k\frac{E_{\mathbf{k}2}}{\omega^2-\xi_{\mathbf{k}2}^{2}}&=&\int\frac{\rho_{2}(E)E}{\omega^2-\Delta^2-E^2}dE\nonumber\\
&=&-\int\frac{\rho_{2}(E)E}{\Delta^2+E^2}(1-\frac{\omega^2}{\Delta^2+E^2})^{-1}\nonumber\\
&\approx&-\int\frac{\rho_{2}(E)E}{\Delta^2+E^2}dE=-\langle\frac{E_{\mathbf{k}2}}{\xi_{\mathbf{k}2}^2}\rangle=\gamma.\nonumber\\
\end{eqnarray}
Here $\rho_{2}(E)$ is the DOS of the active band and Eq. (\ref{gamma}) is derived since $\rho_{2}(E)$ is close to a constant in the vicinity of the Fermi level while away from it, $\Delta^2+E^2\gg\omega^2$.
Finally,
\begin{eqnarray}
\label{p}
\frac{1}{N}\sum_k\frac{\omega(\Delta)}{\omega^2-\xi_{\mathbf{k}2}^{2}}&=&-\pi\rho_{2}(0)\frac{\omega(\Delta)}{\sqrt{\Delta^2-\omega^2}}.
\end{eqnarray}
Combining Eqs. (\ref{G0}) to (\ref{p}), we have
\begin{eqnarray}
\label{p(x)}
p(x)&=&a_3^2-a_1^2x+a_1a_2\frac{x}{\sqrt{1-x}}-\frac{a_2a_4}{\sqrt{1-x}},\nonumber\\
a_1&=&\frac{\eta V\Delta}{2},a_2=V\pi\rho_{2}(0),a_4=\frac{\eta V\Delta_1}{2}\nonumber\\
a_3&=&\sqrt{(\frac{a_2}{2})^2+a_4^2+[\frac{V}{2}(\delta+\gamma)-1]^2}.
\end{eqnarray}
where $x=(\frac{\omega}{\Delta})^2$ and the condition for $p(x)=0$ results in the following equation
\begin{eqnarray}
\label{p(x)=0}
0&=&x^3+(b_2^2-2b_3^2-1)x^2+(2b_3^2+b_3^4-2b_2^2b_4)x\nonumber\\
& &+b_2^2b_4^2-b_3^4,\nonumber\\
b_i&=&\frac{a_i}{a_1},\text{i=1,2,3}.
\end{eqnarray}
If some of the roots of Eq. (\ref{p(x)=0}) lie in $[0,1)$, then IGBS will appear. From above we can see, if $|Q_\mathbf{k}^{11}|^2=0$ and $|Q_\mathbf{k}^{12}|^2=1$, the system can be viewed as a single band model (the active band) with conventional $s$-wave pairing and it is equivalent to setting $\eta=\delta=0$. In this case, $a_1$ and $a_4$ in Eq. (\ref{p(x)}) are zero and $p(x)=a_3^2>0$, which means that a single nonmagnetic impurity will not lead to the IGBS in a single band model with conventional $s$-wave pairing symmetry, consistent with Ref. \onlinecite{zhujx} and references therein.

\begin{figure}
\includegraphics[width=1\linewidth]{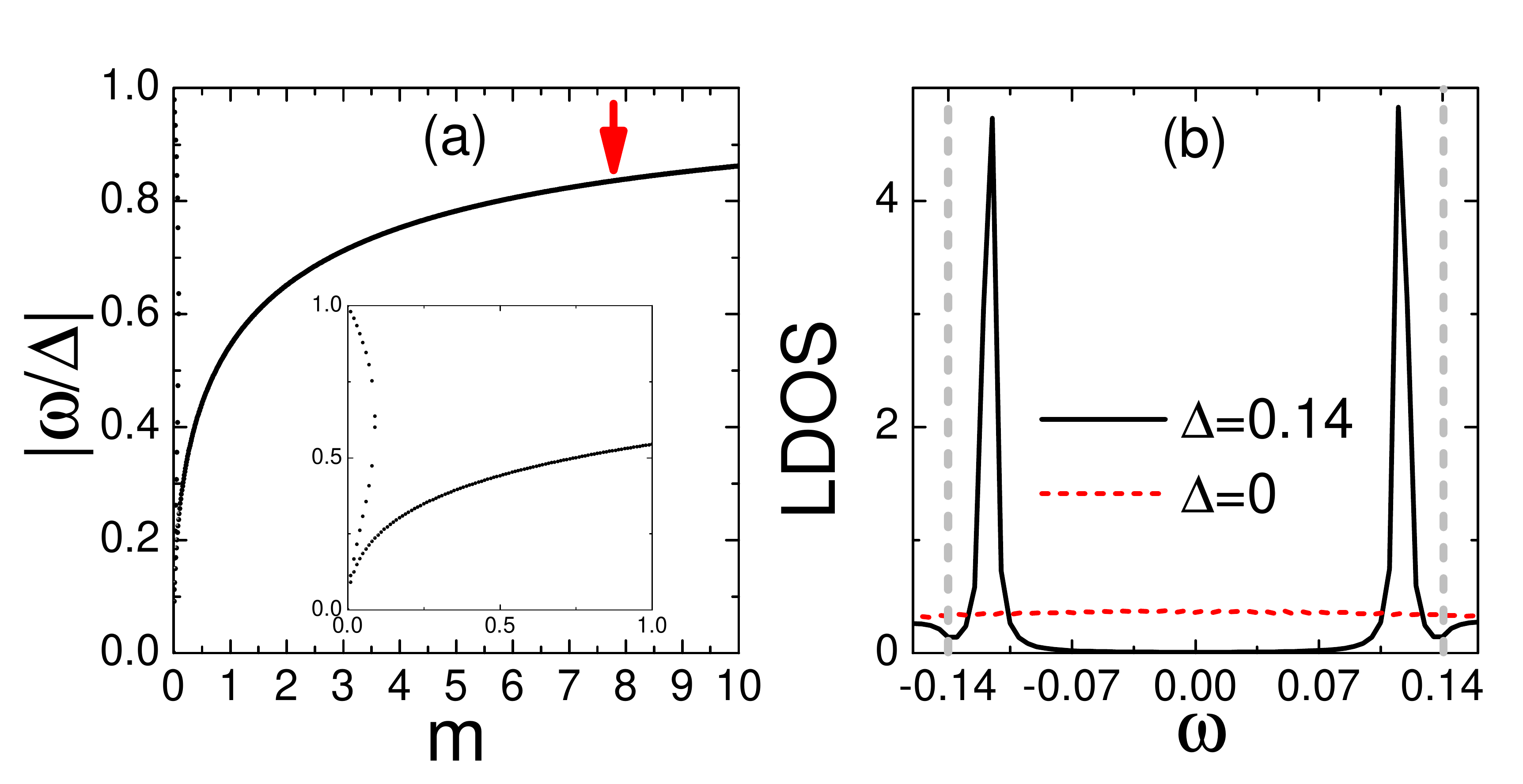}
 \caption{\label{roots} (color online) (a) The solution of Eq. (\ref{case2}). The inset shows the solution in the range $0\leq m\leq1$. (b) The LDOS at the impurity site in the normal (red dashed) and SC (black solid) states, at $V\approx4.74$. The value of $m$ and the location of the IGBS are indicated by the red arrow in (a). The gray dotted lines in (b) denote the SC coherence peaks at $\pm\Delta$. Here $\Delta_1=0$.}
\end{figure}

In the following, in order to illustrate the possible appearance of IGBS with incipient band, firstly we take $\Delta_1=0$. In this case, if $\frac{V}{2}(\delta+\gamma)-1=0$, then Eq. (\ref{p(x)=0}) can be written as
\begin{eqnarray}
\label{case2}
x^3+(2m-1)x^2+m(2+m)x-m^2=0,
\end{eqnarray}
where $m=(\frac{b_2}{2})^2=(\frac{\pi\rho_2(0)}{\Delta\eta})^2$. The roots of Eq. (\ref{case2})
are plotted in Fig. \ref{roots}(a). When $m\gg1$, $|\frac{\omega}{\Delta}|$ will approach 1, which means that no IGBS will exist and the above mentioned single band case ($\eta=0$) corresponds to this situation. In contrast, as $m$ decreases, the IGBS will move deeper inside the gap and multiple pairs of IGBS may even show up if $m$ is small enough [see the inset of Fig. \ref{roots}(a)]. In our model, $m\approx7.8$ and the condition $\frac{V}{2}(\delta+\gamma)-1=0$ can be satisfied at $V\approx4.74$. In this case, the locations of the IGBS should be located at $\frac{\omega}{\Delta}\approx\pm0.83$, as denoted by the red arrow in Fig. \ref{roots}(a). Indeed, the local density of states (LDOS) at the impurity site plotted in Fig. \ref{roots}(b), which is calculated by using the exact expression of Eq. (\ref{G0}), shows two IGBS at $\omega\approx\pm0.115$ ($\frac{\omega}{\Delta}\approx\pm0.82$), agreeing quite well that shown in Fig. \ref{roots}(a) and we can see that the IGBS are indeed induced by superconductivity since they do not exist in the normal state. Furthermore, from the definition of $m$, we can see that $m$ will decrease if $\Delta$ increases, therefore we further set $\Delta=0.2$ to verify the above calculation. In this case $m\approx3.8$, $V\approx4.726$ and the IGBS in the LDOS are located at $\omega\approx\pm0.149$ ($\frac{\omega}{\Delta}\approx\pm0.74$), also agreeing quite well with that shown in Fig. \ref{roots}(a).

If for the active band, there also exists a pairing cutoff $E_c$ [$E_c=E_g=1.245$ in the present model, which is the band edge of the incipient band, as can be seen from Fig. \ref{band}(c)], since we have assumed that the SC pairing does not take place in the incipient band. In this case, the following term should be rewritten as
\begin{eqnarray}
\label{cutoff}
\frac{1}{N}\sum_k\frac{\omega+E_{\mathbf{k}2}}{\omega^2- \xi_{\mathbf{k}2}^{2}}&=&\int_{|E|<E_c}\frac{\rho_2(E)\omega}{\omega^2-E^{2}-\Delta^2}dE\nonumber\\
&+&\int_{|E|<E_c}\frac{\rho_2(E)E}{\omega^2-E^{2}-\Delta^2}dE\nonumber\\
&+&\int_{|E|\geq E_c}\frac{\rho_2(E)}{\omega-E}dE.
\end{eqnarray}
Since $|\omega|<|\Delta|\ll E_c$, the first term on the right-hand side can still be approximated as Eq. (\ref{p}), while the second term is zero since the active band is almost particle-hole symmetric at $|E|<E_c$. In addition, the third term can be absorbed into $\eta$ and $\delta$ since they are all contributed by the normal state Green's function far away from the Fermi level. Therefore Eq. (\ref{p(x)}) is still valid, with slightly varied $\eta$, $\delta$ and $\gamma$. We have numerically verified that the IGBS still exist in this case, similar to those without the pairing cutoff.

\begin{figure}
\includegraphics[width=1\linewidth]{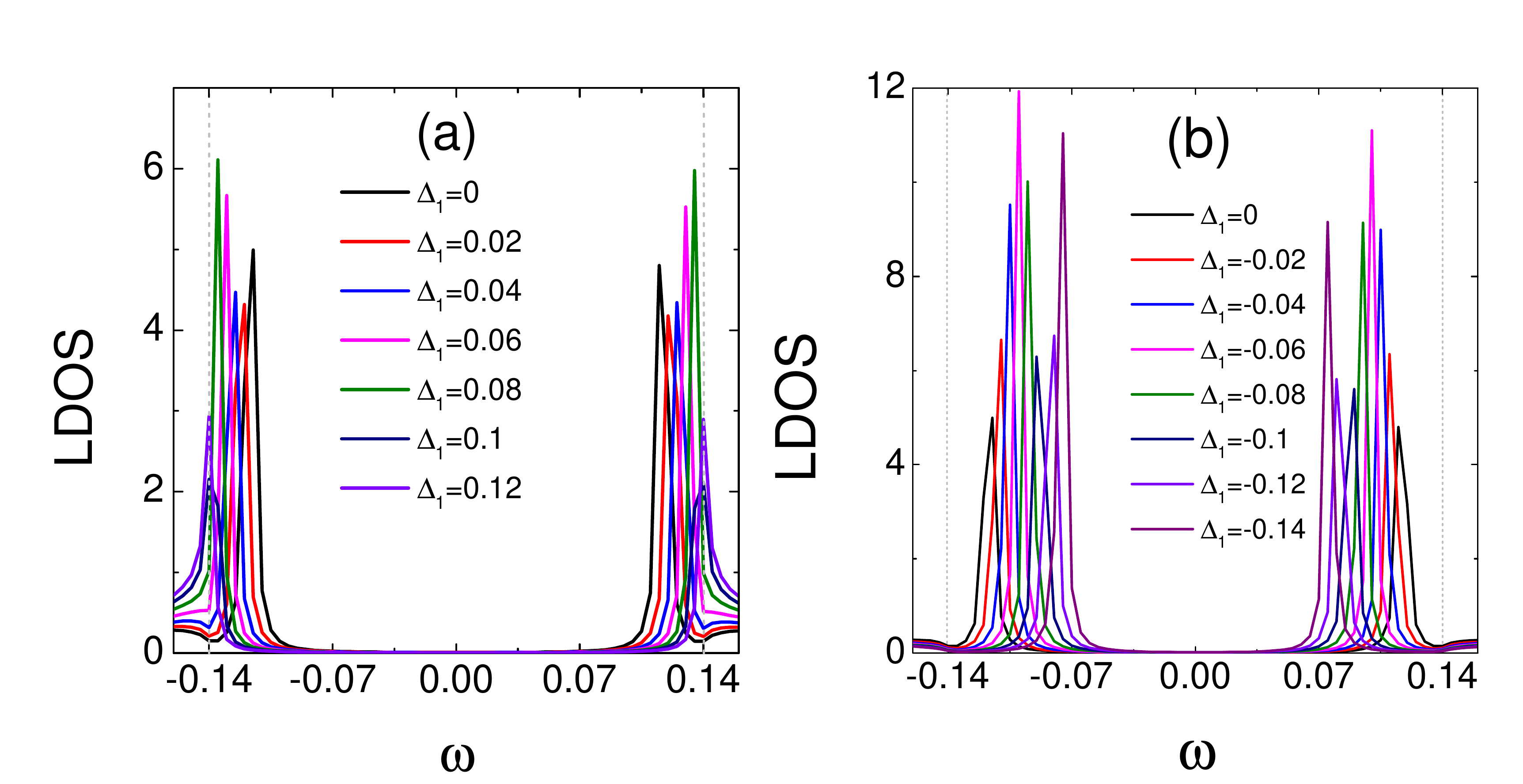}
 \caption{\label{d1} (color online) (a) The LDOS at the impurity site when $\Delta_1$ and $\Delta_2$ have the same sign. (b) is similar to (a), but $\Delta_1$ and $\Delta_2$ have the opposite sign. The gray dotted lines denote the SC coherence peaks at $\pm\Delta$. Here we take $V=4.65$ as an example.}
\end{figure}

Furthermore we have also investigated the effect of $\Delta_1$ on the IGBS. From Eq. (\ref{p(x)=0}) we can see that its roots depend on $\frac{\Delta_1}{\Delta}$ due to the $b_4$ term. If $\Delta_1$ and $\Delta$ have the same sign, the IGBS will move from inside the gap to the gap edges as $\Delta_1$ varies from 0 to $\Delta$ [see Fig. \ref{d1}(a)]. On the contrary, if $\Delta_1$ and $\Delta$ have the opposite sign, the IGBS will move even deeper inside the gap as $\Delta_1$ changes from 0 to $-\Delta$ [see Fig. \ref{d1}(b)]. We have also verified that, the position of the IGBS can well be approximated as the roots of Eq. (\ref{p(x)=0}) as long as $|\Delta_1|,|\Delta_2|\ll E_g$.

Finally we set $\Delta_2=\frac{\Delta_0}{2}(\cos k_x+\cos k_y)$ in Eq. (\ref{unitary}), which is the usual $s_\pm$ pairing function in the 2Fe/cell BZ. Taking $\Delta_0=0.2$ leads the SC gap to vary from $0.159$ to $0.168$ on the Fermi surfaces and the SC coherence peaks are located at $\Delta\approx\pm0.17$. In this case, no matter $\Delta_1=\Delta_2$ or $\Delta_1=0$, IGBS may still show up (for example, at $V=4.65$, the IGBS are located at $\omega\approx\pm0.125$).

In summary, we have investigated the IGBS induced by a single nonmagnetic impurity in sign-preserving $s$-wave superconductors in the presence of incipient bands. We found that, different from the single-band case with conventional $s$-wave pairing symmetry, a single nonmagnetic impurity can induce IGBS in multiband superconductors, even when the SC order parameter maintains the same sign along the Fermi surfaces. Furthermore, the SC order parameter on the incipient bands may affect the position of the IGBS, contrary to the naive expectation. The conclusions seem to be inconsistent with those obtained in Refs. \onlinecite{zhujx2} and \onlinecite{chen}. The reason is, in Ref. \onlinecite{zhujx2}, the incipient bands are at $\sim500$meV below the Fermi level [see Fig. 3 of Ref. \onlinecite{zhujx2}], while in our model and in experiments, it is $\sim100$meV. In Ref. \onlinecite{chen}, the width of the incipient band is $B\geq1$eV (assuming realistic parameters $\Delta_e\approx10$meV and $\Lambda\approx100$meV), while in our model and in experiments, it is $\sim0.2$eV [see Fig. 1(c) of Ref. \onlinecite{wangzf}]. Consequently $\eta$ ($m$) in Refs. \onlinecite{zhujx2} and \onlinecite{chen} is much smaller (larger) compared to that in our model. From Fig. 2(a) we can see that it is such a large $m$ that leads to the disappearance of the IGBS in Refs. \onlinecite{zhujx2} and \onlinecite{chen}. Since the band structure we use agrees better with realistic FeSe-based superconductors, the predictions we made should be more reliable. We have also verified that the conclusions remain qualitatively the same when more realistic two-orbital models of Refs. \onlinecite{gao1} and \onlinecite{gaoy_fese} are used. For example, in Ref. \onlinecite{gaoy_fese}, the formation of the IGBS is numerically investigated based on a two-orbital model and the results are consistent with those in the present work. Therefore the IGBS we studied have to be taken into account when explaining the SC pairing symmetry based on the STM data, especially in the FeSe-based superconductors.

This work is supported by the Natural Science Foundation from Jiangsu
Province of China (Grants No. BK20160094 and No. BK20141441), the NSFC (Grant No. 11374005) and NSF of Shanghai (Grant No. 13ZR1415400). QHW is supported by NSFC (under grant No.11574134).

\end{document}